# Ken Wilson - A Tribute:
## Some recollections and a few thoughts on education

H R Krishnamurthy
Department of Physics, Indian Institute of Science,
Bangalore 560012, India

*Abstract*

*I had the marvelous good fortune to be Ken Wilson's graduate student at the Physics Department, Cornell University, from 1972-1976. In this article, I present some recollections of how this came about, my interactions with Ken, and Cornell during this period; and acknowledge my debt to Ken, and to John Wilkins and Michael Fisher, who I was privileged to have as my main mentors at Cornell. I end with some thoughts on the challenges of reforming education, a subject that was one of Ken's major preoccupations in the second half of his professional life.*

I joined Cornell as a graduate student in physics in the fall of 1972, after completing an M.Sc. degree in physics from the Indian Institute of Technology (IIT), Kanpur [1]. I was pretty much sold on becoming a theoretical physicist, but wasn't very clear what area I would work in. I remember that among the graduate programs I had gained admission to, I was able to eliminate all but Caltech and Cornell relatively easily, but found the final selection between these two a tough one to make. Caltech was the more acclaimed (Feynman and Gell-Mann were among the physics faculty), and had offered me a fellowship; whereas Cornell had offered me only a teaching assistantship. I finally decided to go to Cornell nevertheless, because of the perception that this allowed me to keep my options open for doing either high-energy physics (then more commonly referred to as particle physics) or condensed matter physics, as Cornell was strong in both [2]. Among the flyers I received from the physics department at Cornell, there was one with a list of recent publications from their faculty, and Ken Wilson's name figured prominently at the end, with a string of 4 papers published in 1971, three of them with "renormalization group" in their titles. I looked up the papers in the library, but didn't understand much of what I read, except that the papers seemed very important. By the time I arrived in Cornell, "epsilon expansion" had been discovered [3], and there seemed to be a clear consensus that Ken had achieved a profound breakthrough.

The physics department at Cornell those days had a system that a committee of "four wise men" was designated every year to advise the entering graduate students. Ken was one of these four the year I joined, and by a fantastic stroke of luck, I was assigned to him. I met him soon after my arrival at Ithaca, and asked him whether I could skip the first year graduate courses in Quantum Mechanics (QM) as I felt I knew the course material well. He said he would give me a written exam, and advise me based on my performance. The exam consisted of a bunch of QM problems, and I had no difficulty solving them. When I met him later, after he had a chance to look at my solutions, to my great joy he suggested that I skip *all* the first year graduate courses, and instead credit the second year courses, including a special topic course that he was teaching that fall term on the renormalization group (RG) and epsilon expansion (from notes that were later published as the seminal Wilson-Kogut *Physics Reports* article [4]). I did as he suggested, but partly because of my limited exposure to advanced statistical physics, field theory and critical phenomena, and partly because of limitations in my approach to learning (I was very reluctant to plunge in and learn something without having mastered what I considered the "prerequisites"), found his course rather difficult to cope with. I eventually dropped out of crediting the course (though I probably continued to sit in on the lectures), and focused on the other courses I was taking, on Solid State Physics [5] and Quantum Field Theory. I don't think I had much interaction with Ken the rest of that academic year, except for occasional meetings for him to sign papers as my adviser.

During my second semester at Cornell, and during the summer that followed, I tried my hand a bit at experimental physics, by doing a couple of projects with Bob Buhrman, which also helped me to fulfill the experimental physics course requirements that were mandatory for all physics graduate students at Cornell. Bob was a wonderful person to work with, but by the end of this period it was clear to me that I should stick to theory. I was also clear that I wanted to work on problems related to RG and critical phenomena. I met Ken and asked him whether I could work with him. He basically said yes, but added that he had switched his interest to lattice gauge theory [6], and asked whether I would be interested in working in that area. I told him that I was keener to work in condensed matter theory, and he suggested that I should talk to Michael Fisher. So I went and met Michael, and joined his group, moving to a desk in Baker Hall. By then David Nelson, a whiz-kid who was at that time in his $4^{th}$ year of Cornell's famous "6 year Ph.D program" [7], had already started working with Michael on projects in RG. Michael was perhaps a bit hesitant to start me on similar problems, and asked me to look at a few other problems that were of interest to him, in more traditional areas of statistical mechanics. I did that, but couldn't get myself interested in those problems, and was puttering around for a bit, unclear where I was headed.

Then, one momentous day, to my utter amazement, the trio of Ken Wilson, Michael Fisher and John Wilkins sought me out, I think after a departmental colloquium. John was away on a sabbatical my first year at Cornell, so I had actually not met him until that moment, but had only heard of him. The first thing he said to me after being introduced was, "So you are the sucker"! It turned out that they had in mind a proposal for a research project for me to take on. A little prior to that time, Ken had completed the invention of a remarkable new technique, the numerical renormalization group (NRG), and used that to achieve yet another breakthrough – the solution of the celebrated "Kondo Problem" connected with magnetic impurities in metals [8]. The solution had been announced in a Nobel Symposium in June 1973 in Göteborg, Sweden [9], but the details were not yet published. The proposal was to have me explore the extension of the technique to the more microscopic Anderson Impurity Model (AIM) [8]. I took a bit of time to do some browsing of the literature on Quantum Impurity Problems, and of Ken's notes on the NRG (the notes were later expanded and included in Ken's second famous review article on the RG, in *Reviews of Modern Physics* (*RMP*) [10]), found myself hooked, and signed on to the project

[11]. Ken and John became the co-advisers of my thesis research, and Ken, John, Michael and Bob the members of my special committee.

My next two years were the most intense and wonderful period of learning I have ever experienced. I had the privilege of learning the intricacies of the renormalization group, in particular, the numerical renormalization group, from its creator, Ken. In fact, Ken gave a whole new set of lectures, essentially on the material that later went into his *RMP* article [10], and this time, I was ready; I found his lectures to be amazingly clear and insightful, and soaked them up. John was very generous in sharing his expertise and time to help me learn quantum impurity physics. I learnt a whole lot about RG and critical phenomena from Michael and his group, especially his group seminars. There were very many other exciting things happening at Cornell as well during this period, on which there were lectures and seminars I could learn from [12]. In addition to all this, I honed my computing skills developing my own NRG code for solving the AIM, learning by example from Ken, a master programmer.

Actually, the additional coding required to extend the NRG to the Anderson Impurity Model compared to the Kondo problem is relatively minor. So Ken suggested initially that I modify the NRG program he had written for the Kondo problem and use that for the AIM project, and gave me a copy of his program. I was flabbergasted when I saw it – it had well over a thousand lines of code, pretty much as one single program (except for calls to a matrix diagonalization subroutine), and there was not a single comment statement in it! Many important variable names were chosen in ways I could not fathom; there was an XXXX and a YYYY! I had to go through the code line by line, annotating it along the way, which took me a while; then I understood and appreciated how tightly and intricately knit it was. All available symmetries of the Hamiltonian had been used to reduce the sizes of matrices to be diagonalized to the minimum possible, and storage of arrays had been maximally optimized to reduce memory requirements. I have always wondered how Ken kept track of what was what in the program, and how he debugged it. Knowing how awesome he was as a programmer (he was one of the very few physicists I have come across who knew how to write machine code, and would use it to optimize the innermost computations inside 'do loops'), I am inclined to believe that he had such algorithmic clarity that he wrote code that needed very little iteration and debugging; and that he had prodigious memory which helped him keep track of obscure variable names! I wish I had preserved a copy of Ken's program for posterity, but unfortunately did not have the foresight to do so.

In any case, being more of a mortal, I was terrified by the thought of modifying his program, having it bomb, and being unable to debug it. So I decided to write my own program starting from scratch. I probably spent too much time writing it in a modular fashion, putting in lots of comments, choosing variable names carefully so that they corresponded as closely as possible to the physical quantities they represented, and so on; but it was a great learning experience, and Ken and John were generous in allowing me this leeway. After benchmarking my program to ensure that it reproduced Ken's results for the Kondo problem, I did eventually start producing exciting new results for the AIM, and had great fun analyzing them and thinking about the physics that they represented. The work on the symmetric Anderson model, plus a comprehensive review of the literature, constituted my thesis [13]. The process of completion of the other "formalities" connected with my getting the Ph.D. degree was incredible for its informality. I left for a post-doc at the University of Illinois at Urbana-Champaign (UIUC) in August 1976, after having deposited the handwritten manuscript of my thesis with the venerable Velma Ray, who was Hans Bethe's secretary. She was legendary for her skills in typesetting theses with lots of equations, and I certainly needed her to typeset mine. The understanding was that I would come back by the end of that year for my thesis defense, but the process was actually completed only in Jan 1978 [14]! I have forgotten the details about how this came about - perhaps everybody involved, including me, forgot that my thesis defense was not yet a done deal, until more than a year had gone by!

My periodic interactions with Ken regarding my thesis research were invariably rather brief, but pleasant and rewarding. Ken was very informal - I never had to make an appointment to see him, and would walk into his office whenever I wanted to, which was typically when I had some progress to report, or to seek help when I faced some obstacles in my work. There I would generally find him, mostly in his signature grey pants and white shirt, often with his feet up on the table, and deep in thought. But he never seemed to be perturbed by the interruption, and would turn to me with the twinkle in his eye that used to be a ubiquitous feature of his demeanor, as can be seen in so many of his photographs [15]. When I reported to him the newer aspects of the physics of the AIM as they began to emerge from my NRG calculations, which I thought I was the first to discover, he would most often just nod in agreement, and it was somewhat disconcerting to find that they seemed obvious to him!  When I ran into an obstacle, and broached it to him, there were only two possible outcomes - if he had a hunch as to how one might be able to get around the obstacle, he would say it succinctly, and his hunches almost always helped out; otherwise, he would say he hadn't thought about the issue, and did not have any comments that might be of help. So it was somewhat difficult for me to hang around in his office for long - he never seemed to be one for "small talk". He was also probably too absorbed in his work on lattice gauge theories during this period to be very actively involved in what I was doing. I remember though, that sometime after we published the first NRG results on the symmetric Anderson model [16], he asked me how it was being received by the condensed matter community. When I said I thought it was being well received [17], he did seem pleased. I don't know whether he kept track of the fact that the NRG has continued to thrive, especially as a solver for quantum impurity problems that arise in the context of quantum dots and Dynamical Mean Field Theory [18]; if he did, his pleasure would have been even greater.

At Cornell I also had some opportunities to observe closely Ken's approach to teaching. I was his teaching assistant for an undergraduate course on electromagnetism that he taught one of the terms, and I sat in on most of his lectures. Purcell's book was the text.  My memories of the details of this experience are a bit hazy, but the impression I have retained is that Ken was not very particular about sticking to the textbook material or of covering a pre-planned set of topics, but would spend time on what he thought was interesting and useful. Even in a well-worn subject like electromagnetism, I found many of his comments and observations very original and insightful – but they might have been lost on many of the students if they did not have prior exposure to the material. He tended especially to emphasize numerical techniques of solving electromagnetics problems. For example, he taught the students the numerical technique for solving boundary value problems involving the Laplace equation for the electric potential, by choosing a square or cubic grid of points, and iteratively updating the potential at each site to the average over all its nearest neighbor sites, in great detail. This was very much in keeping with his abiding interest in computers and computing [19], and his vision and foresight that computers were going to play a major role in the future of physics, and it was important that students get an early exposure to numerical techniques. Ken followed up on his vision in many ways. He was an active campaigner for improving computer resources for research [20]. He headed an initiative that he christened the "Gibbs project" [21], which attempted to create what he thought would be the ideal computing environment for physicists. What he was visualizing was *one resource* that combined the best features we have come to see in MAPLE, MATLAB, Mathematica and program libraries such as LAPACK, etc., and was user-friendly enough not to require the learning of a new language. I think we are still rather far away from having anything like what he visualized.

Unfortunately, I did not keep in regular touch with Ken after my return to the Indian Institute of Science in Bangalore, India in 1978. This was difficult to do in the initial years in any case, as it could have been done only by snail mail, and I was a bit hesitant to write letters to him and expect him to reply. I was overjoyed by the award of the Physics Nobel Prize to him in 1982, of course, did write him a congratulatory note, and I think I did get a brief response. After the advent of e-mail, I would send him Christmas and New Year greetings occasionally, and he would not always respond. But I got to meet him briefly now and then, during visits I made to Cornell, and later to the Ohio State University [OSU] at Columbus, Ohio, where both John Wilkins and he moved in 1988.

As I look back, I am astonished as to how many of Ken's values [19] I seem to have imbibed, some consciously, and many subconsciously, that have heavily influenced me in my career and life. For example, in the context of his work on the Kondo problem, Ken had carried out some extremely tedious $4^{th}$ order (Rayleigh-Schrodinger) perturbation theory calculations for the energy levels of the Kondo NRG Hamiltonian – pages after pages of neat algebra, meticulously listing out the various terms that arise. In the course of my own research work, if I felt discouraged by some tedious and daunting algebra that needed to be done for me to make further progress, I could draw strength from his example – if a genius like Ken could sit down and carry out tedious algebra, surely I had to discipline myself to 'just do it'! I also found his commitment to societal reforms inspiring. The most prominent, and well known, are of course his role in making supercomputing widely available to the research community [20], and his involvement in reforming education [22].

Reforming education was the major concern and enduring passion of Ken in the second half of his career, and perhaps the main factor responsible for his move from Cornell to OSU in 1988. A major landmark in his work on education was the publication of the book "Redesigning Education" [23]. A summary of his contributions and leadership role in reforming education in the US, and references to some of the articles he wrote on education, can be found in the obituary issued by the OSU [22].

Most of the articles in this memorial volume on Ken are related to his research contributions in physics. The impact of his work in physics is certainly more widely recognized than his work in education, due in no small measure, of course, to the award of his Nobel Prize. Indeed there might be many who feel that he wasted his talents working on such a complex, perhaps insoluble, problem as education. However, in the context of how Ken himself saw his role in physics research vis-a-vis his role in education, I remember him saying something along the following lines (unfortunately I have not been able to locate a precise quotation that I can reference), which I have always found heartening. While in physics the importance of a contribution often depends on a problem being *solved*, in case of societal issues such as education, even if one is able to contribute to a 1% improvement by some measure, (for example, the fraction of students getting education better than some threshold), it can make a huge difference to a very large number of people! Hence I thought it might be fitting to end my tribute to Ken by airing some thoughts on the challenges confronting educational reform, especially in India.

The huge challenges that confront India in particular, and the developing and developed world in general, as regards educational reform at all levels, i.e., school, college and vocational, are of course well known. India is now entering a period of "demographic dividend" – the period when the growth rate of the working age population well exceeds the growth rate of the overall population [24]. However, as eloquently expressed by Nandan Nilekani [25], preeminent cofounder of the Indian IT company Infosys, for this to be truly a "dividend", the working age population needs to be productive. Hence, providing them education, at the least vocational education to impart to them productive skills, is an obvious necessity. If this is not available, we will instead be confronted by a "demographic disaster". The numbers are staggering - around 64% of India's population (of well over a billion) is expected to be in the age bracket of 15–59 years by 2026 [24].

National agencies in most countries are very much engaged in confronting the challenges of education, and have set many commendable goals [26]. However, numerous widely recognized obstacles block their way forward in achieving these.

A major one of these, especially in India, is the dreadful shortage of trained, high quality teachers at all levels. There is also a shortage of teaching material and laboratory and other resources. What is worse, even when these are available, the 'education' imparted is not designed to suit the people in need of the education. Conventional methods of imparting education, based on lectures, rigid curricula, undue emphasis on performance in centralized examinations rather than on learning, and their use for filtering students at various levels, governed by the "one size fits all" paradigm, are widely prevalent in India. Their limitations are well known and well documented; and they are woefully inadequate for the challenges at hand given the diversity of the student population, leaving large sections of them poorly educated and disheartened. Many voluntary groups and organizations have made laudable efforts addressing these among small batches of students all over India, but the major challenge is to come up with solutions that are *scalable to the huge numbers* of people that need the education.

Current breakthroughs in computer and communication technology, especially mobile devices, and the widespread and rapid increase in their accessibility in all parts of India, open up immense new possibilities, and perhaps a new paradigm, both for "*tailor making*" education to suit individual and societal needs, and for addressing the problem of *scale*. I discuss below some key ingredients which I believe are urgently needed for this, in the context of three clearly distinguishable aspects of education: (1) identification and publicizing of what needs to be taught or learnt, (2) teaching and learning and (3) testing and evaluation of proficiency.

Imagine that groups of people come together, perhaps supported by philanthropic foundations, and perhaps in cooperation with governments, to take the lead in the creation of a novel, model "learning tree": a framework of creatively designed, hierarchical but interconnected modules of learning content, covering all levels, from primary through college, and all aspects of human endeavor, including the learning of life skills. This model content is to be designed so as to make the best use of the wide variety of formats and types of learning material and learning methods that are viable, such as video lectures and presentations, audio lectures, movies, multimedia presentations, computer simulations, lab work, field work, computer games, mini projects, etc., in such a way that it can be flexibly adapted to educate students with varied backgrounds, levels and requirements. The primary design constraint is that students should be able to learn the material, either individually or in small groups, by working through them at their own pace, assisted by teachers who act as mentors or coaches rather than as lecturers and graders. The creation of such learning modules can surely be accomplished by teams of highly qualified and committed educationists with expertise in the different areas and people with skills in multimedia content creation working together.

Imagine that, in synergy with such a learning tree, a "testing tree", of hierarchical and interconnected testing modules, is created. Each testing module is to be associated with one or more learning modules, and consist of very large banks of carefully designed questions, problems and other testing methods, of graded difficulty levels, and covering all the well-recognized objectives of learning [27]. The question banks need to be made large enough that random selections from them can be used by the students to evaluate their own progress *while they are learning*, as well as for the purpose of final testing and certification of the extent to which they have mastered the content of a learning module. The design goal, and the challenge, is to ensure that the *evaluation can largely be done by computers*, with an end result that is nevertheless an objective, unambiguous, tamper-proof evaluation of the proficiency level the students have attained in that particular learning module [28].

In recent years there has been a phenomenal growth in the availability of open source learning content of various types and levels, and of steadily rising quality [29]. But a large fraction of it consists of conventional lecture courses, slide presentations and texts. We have a long way to go before anything along the lines envisioned above becomes available. Picture a gigantic network, with model modules of learning content at every node, each connected hierarchically with numerous other nodes, which students can traverse along their own paths, and at their own pace… I believe that such a framework still needs to be designed and built, and the existing and upcoming open source learning content can then be hyper-linked to the model learning modules.

As regards evaluation content, my impression is that what is currently available in open source is very limited, both in quantity and quality. Some proprietary learning and evaluation content of fairly high quality is probably available, but the majority of students in India, for example, cannot afford the costs for getting access to these. I am sure many people will be skeptical that such testing modules as I am envisaging above can ever be created in open source, and even if that is done, that they will be viable as sure means of evaluation. I am inspired, however, by the shining examples of the creation of open source software such as LINUX and GNU, and of Wikipedia. I believe that if a large enough group of us are enrolled into thinking of this as a desirable goal, and put our creative energies to work, we can achieve it [30].

The amount of work and the challenges involved in developing such learning content and testing modules are huge, but the payoff is enormous. Once the framework is designed and created, we will have a resource that we can continually improve and expand, based on user feedback, as well as new developments and discoveries. Somewhere along the way, when a consensus begins to emerge that the learning modules are effective in helping students to learn, and that the testing modules are indeed objective evaluators of the test taker's proficiency level in the associated content, we will be poised for a paradigm shift in the education system.

For example, we can do away with the "one size fits all" education system, with rigid schedules and time deadlines for all students to attain proficiencies in specific courses at the same pace. What is then likely to happen is that, depending on their backgrounds and abilities, and the availability of mentors, students will take different amounts of time to master a module at specific levels of proficiency. But when they do, the mastery is standardized. When the students complete the learning of a module, they will have many choices: they can put in more effort to improve their proficiency level in the same module if there is room, or go on to a higher level module in the same sub-area, or a different module in the same subject area, or a different module in a different subject area, based on the prerequisites built into the learning modules. It is likely that students will choose to attain different combinations of proficiency levels in modules and subject areas, according to their own individual preferences, abilities and career goals.

The availability of standardized and graded learning content and proficiency tests opens up the possibilities for students who have attained the required proficiency levels in any subject area to themselves act as mentors for the students who are learning appropriate modules that are a few levels lower. People who are already in the teaching profession, but have been inadequately trained, can also use the same learning modules to improve their proficiency levels and become more effective teachers. I see this as a positive feedback process that can eventually eliminate the bottleneck of the shortage of qualified and competent teachers. The opportunities that it creates for the retraining of people who are already employed as the need arises are obvious.

The above paradigm of education also helps to limit the large scale branding of students based on performances in centralized qualifying examinations that is prevalent, which is very stressful for the students. In India, coaching centers that coach students to cram for such centralized examinations are all pervasive. If the testing modules are such that it is impossible for anyone to be coached to do well in the tests without having actually mastered the learning modules, the coaching centers will be forced to turn into education centers. Employers can hire people based on the proficiency levels needed by them, with full confidence that the employees have actually learnt what they are certified to have, and without having to invest in retraining the students as they need to do at present. Students with disadvantaged backgrounds can reach the same proficiency levels as those without, simply by being provided with the required time and assistance in learning – I see this as true affirmative action.

In summary, I see the creation of publicly available, primarily web based, modular learning and evaluation content of the sort envisioned above as the key to cracking the problem of *scale and quality* in education; this especially so in developing countries like India where proprietary content is unaffordable to the majority of the students that need to be educated. The arena of teaching or learning of the publicized or equivalent content will then be open for the dance of human enterprise. Just imagine what the pace and extent of humanity's progress can be when the number of well-educated, productive and creative people is in the billions!

*Acknowledgments*

*The seeds for the opportunities that made it possible for me to get to Cornell were sown in my upbringing, in the encouragement for learning that I received from my parents, both schoolteachers. Several excellent teachers in my school and undergraduate college helped as well. But the clincher, as I have stated [1], was my stint at IIT Kanpur, where H. S. Mani and T. V. Ramakrishnan, in addition to being great mentors, encouraged me to pick Cornell over Caltech. To all of these people, I owe a lot.*

*Much to my chagrin, my administrative commitments as the Chair of the Physics Department at the Indian Institute of Science (IISc) (from 2010-14) made it very difficult for me to keep my promise to contribute to this memorial volume within the original time schedule, and I had actually given up on the idea. However, Belal Baaquie, my classmate at Cornell, co-student of Ken, and co-editor of this memorial volume, would not give up on me. My stepping down as chair and coming away on a sabbatical leave to the University of California Santa Cruz (UCSC) opened up the possibility again. I thank Belal and K K Phua for their patience and encouragement, and Sriram Shastry, my host, and the Physics Department at UCSC for their enabling support. I also thank Sriram for reading through a draft of the article and making valuable suggestions for improvement.*

*The bouts of intense concentration required for completing such an article meant that I was not available for many things at home when I was engaged in the writing. I thank my wife, Raj for her patience and support. She and my son Chaitanya also read through a draft, and pointed out many typos and sentences that could use improvement.*

*The thoughts on education I have put down here have been evolving over a period of time. They are certainly influenced by my experiences in teaching physics over the past 40 plus years, the many articles and books on education that I have read, and the many discussions I have had with colleagues in IISc and elsewhere, but in ways that are difficult for me to acknowledge specifically. I had occasion to write down some of these thoughts for a draft of an "educational technology initiative" proposal for the International Centre for Theoretical Sciences (ICTS - see http://www.icts.res.in/home/) some years ago, for which I have to thank Spenta Wadia and Avinash Dhar. My thoughts have been sharpened by my recent experiences as the Physics coordinator of the new B.S. program we started four years ago at the IISc, and as an instructor for the "Thermal and Modern Physics" course that is taught in the 3$^{rd}$ semester of this program. In particular, for this course we experimented with supplementing traditional lectures and lab training with computer based adaptive learning, using open course content, and an adaptive learning platform created by an educational technology startup "Lrnr Adaptive Learning Solutions (Pochys Ventures, Inc.)" (website: lrnr.us). I thank Aravind Pochiraju of lrnr for many discussions about education, and for educating me about Blooms taxonomy.*

vastly more extensive, creatively crafted, modular versions of these that are also available in the public domain to students as self-evaluation tools *while* they are learning.
29. E.g., coursera (https://www.coursera.org/), edx (https://www.edx.org/), Khan academy (https://www.khanacademy.org/), Tedtalks (http://www.ted.com/) …. Many top ranking universities, such as MIT, Yale, Stanford, … have created and put out open courseware. California State University's MERLOT collection (http://www.merlot.org/merlot/index.htm) provides links to as many as 40000 free open course materials, 5000 free online courses and 3300 free e-textbooks.
30. From numerous conversations I have had with colleagues in India as well as elsewhere over the years, it is my impression that while most of us who are in the teaching profession like the mentoring aspects of teaching, we find it a chore to set and grade assignments and exams. Furthermore, the evaluation is non-standard, grade inflation is pervasive, and the assigned grades are often not true measures of the proficiency attained. If we embrace the creation of open source testing modules, we can transform the process of evaluation into a collective creative enterprise, and take the chore out of teaching.